\documentclass{aa}  
\usepackage{graphicx}
\usepackage{hyperref}
\usepackage{txfonts}
\newcommand{\HI}{\ion{H}{i}}
\def\kms{km\,s$^{-1}$}

\usepackage{pdfpages}
\usepackage{graphicx,graphics,color}

\usepackage{ulem}

\begin{document}

   \title{uGMRT HI 21-cm absorption observations of two extremely inverted spectrum sources}

\author{Mukul Mhaskey \inst{1,2}\fnmsep\thanks{mukul@tls-tautenburg.de}, Surajit Paul \inst{2}, Neeraj Gupta \inst{3}, Dipanjan Mukherjee \inst{3}  
          \and
          Gopal-Krishna\inst{4,5}
          }

   \institute{Th\"uringer Landessternwarte, Sternwarte 5, D-07778 Tautenburg, Germany
%              \email{mukul@tls-tautenburg.de}
         \and
             Department of Physics, Savitribai Phule Pune University, Ganeshkhind, Pune 411007, India
             %\email{surajit@physics.unipune.ac.in }
         \and
             Inter University Centre for Astronomy and Astrophysics (IUCAA), Pune 411007, India
             %\email{ngupta@iucaa.in, dipanjan@iucaa.in}
          \and
             UM-DAE Centre for Excellence in Basic Sciences (CEBS), Univ. of Mumbai Campus, Vidya Nagari, Mumbai 400098, India
           \and
             ARIES, Manora Peak, Nainital 263002, Uttarakhand, India
            }
   \date{Received May xx, xxxx; accepted xxxx xx, xxxx}

\abstract{We report the detection of HI 21-cm absorption in a member of the rare and recently discovered class of compact radio sources, \lq\lq Extremely Inverted Spectrum Extragalactic Radio Sources (EISERS)\rq\rq~. EISERS conceivably form a special sub-class of the inverted spectrum radio galaxies since the spectral index of the optically thick part of the spectrum for these sources crosses the synchrotron self absorption limit of $\alpha=+2.5$ (S($\nu$) $\propto$ $\nu^{\alpha}$). We have searched for HI absorption in two EISERS using the recently upgraded Giant Metrewave Radio Telescope (uGMRT)  and detected an absorption feature in one of them. The strong associated HI absorption detected against the source J1209$-$2032 ($z$=0.4040) implies an optical depth of 0.178$\pm$0.02 corresponding to an HI column density of 34.8$\pm$2.9 $\times$10$^{20}$ cm$^{-2}$, for an assumed HI spin temperature of 100 K and covering factor of 1. This is among the highest known optical depth and HI column densities found for compact radio sources of GPC/CSS type and supports the free-free absorption model for the steeply inverted radio spectrum of this source.
For the other source, J1549$+$5038 ($z$ = 2.171), no HI absorption was detected in our observations.}


\titlerunning{uGMRT HI 21cm absorption observations of EISERS}
\authorrunning{Mukul Mhaskey et al}
   \maketitle
%
%-------------------------------------------------------------------
\section{Introduction}
The gigahertz-peaked-spectrum sources (GPS) and compact-steep-spectrum sources (CSS) which are known to be evolving young radio galaxies are ideal candidates to study the co-evolution of radio jets and their host galaxies \citep{O'dea1998}. The inverted spectrum in the GPS and CSS sources is a result of absorption of synchrotron radiation at low (sub-GHz) radio frequencies, either due to synchrotron self-absorption (SSA) and/or free-free absorption (FFA) processes \citep[e.g.][]{O'dea1998}. The GPS/CSS sources have small physical extent ($<$10 kpc) but morphologies similar to 
large-scale ($>$100 kpc) powerful radio galaxies with giant radio lobes \citep{Wilkinson1994, Stanghellini1997, Orienti2006}. The youth scenario for these sources is well supported by  the estimates of age ($<$ 10$^{5}$ years) coming from the proper motion of their hotspots obtained from very long-baseline interferometry (VLBI) and, indirectly, from spectral ageing analyses \citep{Owsianik1998, Polatidis2003, Gugliucci2005}.

The radio spectrum predicted from numerical simulations of jet-ISM interactions in young AGNs have shown that the slope of the inverted part of the low radio-frequency spectrum depends on the age of the radio source, with the most inverted spectral index corresponding to the youngest sources \citep[see Figure 1,][]{Bicknell2018}. Hence, in an attempt to understand the evolution of AGNs and study the absorption mechanisms operating in such sources, we started a program using the upgraded Giant Metrewave Radio Telescope \citep[uGMRT,][]{Gupta2017}, to identify inverted spectrum sources that have a spectral index, $\alpha$ (between 150 and 325 MHz, observed quasi-simultaneously) greater than the critical value of $\alpha$ $>$ $+$2.5, the theoretically allowed spectral index for SSA occurring in a homogeneous radio source \citep{Slish1963,Kellermann1966}. The program has led to detection of seven sources with $\alpha$ $>$ $+$2.5 (within uncertainties), four out of them having  $\alpha$ $>$ $+$2.75. These sources were termed as \lq Extremely Inverted Spectrum Extragalactic Radio Sources (EISERS)\rq \citep{GopalKrishna2014,Mhaskey2019a,Mhaskey2019b}. The numerical fitting based on different absorption models, to the observed spectral energy distributions for the seven EISERS indicates that the FFA model within external in-homogeneous screen of thermal plasma \citep{Bicknell1997,Bicknell2018} can be the preferred alternative, since fitting the SSA model would need a departure from the standard synchrotron emission scenario which is based on a {\it power-law} energy distribution of the radiating relativistic electrons \citep[][]{Rees1967, Begelman1996,GopalKrishna2014}.

The FFA model predicts free-free absorption of the synchrotron emission from the radio source by the intervening ISM of the host galaxy, likely in the vicinity of the radio source. Quite plausibly, the efficiency of the FFA is strongly dependent on the presence of dense ionised gas in the circumnuclear region \citep{Peck1999,Kameno2000,Marr2001}. Indications for such dense plasma regions could come from a variety of muti-wavelength studies, such as the detection of broadened emission-lines, indicating a strong interaction between the radio jets/lobes and the ISM \citep{Gelderman1994, Zovaro2019}. Likewise, the discovery of high rotation measure and indications of strong depolarisation, would indicate the presence of high electron densities and magnetic fields \citep{Stanghellini1998}.

HI observations provide an alternative means to study the gas content in the GPS/CSS sources. Although, only a minor fraction of the total gas content in the ISM comprises of cold neutral gas, it provides a lower limit to the total gas mass and density \citep{Pihlstrom2003}. Additionally, the characteristics of the HI absorption profile can provide valuable information on the gas kinematics within the narrow-line-region (NLR) of the host galaxy and ascertain whether a localised outflow of neutral gas is occurring due to jet-ISM interactions, as predicted by some recent simulations \citep{Bicknell2018, Mukherjee2016} and indeed observed in neutral, molecular and ionised gas components likely associated with the hosts of some radio galaxies \citep{Morganti2018}.

An understanding of the distribution and kinematics of different components of the circumnuclear gas and the extended ISM in the host galaxy is essential for understanding the mechanism fuelling the radio activity, the evolution of the radio source properties within the framework of the unified scheme. Over the years, HI 21-cm absorption line studies of radio sources have emerged as a powerful tool to constrain the properties of cool atomic gas associated with AGNs \citep[e.g.][]{Morganti2001, Vermeulen2003,Gupta2006b, Gereb2014, Aditya2018}. Moreover, a difference in properties of the HI absorption in extended and compact radio sources has been reported \citep{Morganti2018}, such that the absorption is detected more often in the case of compact (GPS/CSS) sources- i.e., young radio sources still evolving through the host galaxy 's ISM, as compared to the extended radio sources. The detection rate of HI absorption is found to be as high as 40-45\% \citep{Morganti2018}. On the other hand, \citet{curran2010} found that the 21cm absorption is not detected in objects which have a high ultraviolet luminosity (L$_{UV}$ $\sim$ 10$^{23}$ W Hz$^{-1}$. Hence they propose that the the presence of 21 cm absorption shows a preference for radio galaxies over quasars and the higher detection rate in compact sources (e.g. CSS/GPS) may be biased by inclusion of sources with high ultraviolet luminosity. 

With the above motivation, we have recently conducted a HI absorption study of two EISERS, out of the total seven EISERS mentioned above, since the spectroscopic redshifts for the two could be obtained from the literature. In this paper we report the results from this HI study and place them in the context of similar detections reported for GPS/CSS radio sources. The paper is divided into five sections. In Section 2 we describe the sample and its multiwavelength observations (optical and VLBA) taken from the literature. Details of our uGMRT HI observations and analysis are reported in Section 3. Section 4 contains a brief discussion of the observational results, followed by a summary of this study in Section 5. 

\section{Sample of EISERS for HI absorption}
The two sources, J1209$-$2032 and J1549$+$5038 for which spectral line observations are reported in this paper belong to the sample of seven EISERS having a spectral index, $\alpha$ $>$ $+$2.5 between the frequencies 150 and 325 MHz, as found by conducting quasi-simultaneous uGMRT observations at the two frequencies. The two sources were selected because their spectroscopic redshifts are available in the literature, which is a critical requirement for HI absorption spectroscopy. For the remaining five sources, unfortunately this crucial information is lacking. In the future, we intend to start a separate program to measure the redshifts of EISERS, so that the sample may be enlarged to arrive at general conclusions. The coordinates, optical identifications and flux densities for the two sources, namely J1209$-$2032 and J1549$+$5038, are provided in Table~\ref{table:spec-prop}.

\subsection{J1209$-$2032 and J1549$+$5038}
\begin{itemize}
    \item \textit{Radio properties}: \\
    The sources J1209$-$2032 and J1549$+$5038 are inverted spectrum source with spectra peaking near the rest-frame frequencies of 430 MHz and 1407 MHz, respectively. The spectral index (150-325 MHz) below the turnover frequency is $\alpha >$ +2.64 for J1209$-$2032 and $+2.42\pm$0.22 for J1549$+$5038 \citep{Mhaskey2019a,Mhaskey2019b}, whereas the  theoretical SSA limit for a homogeneous synchrotron radio source is $+ 2.5$. The AT20G catalogue reports fractional polarisation of 1\% at 20 GHz, 8 GHz and 5 GHz, for J1209$-$2032 \citep{Murphy2010}. For the source J1549$+$5038 \citet{Pasetto2016} have inferred a large (rest-frame) rotation measure (RM $=$ 1400$\pm$500 rad m$^{-2}$). 
    The available radio continuum data for J1549$+$5038 confirms a strong flux variability at centimetre wavelengths, as cited in \citet{Mhaskey2019b}. 
\end{itemize}

\begin{itemize}
    \item \textit{VLBI structure}: \\
    The milli-arcsecond resolution images of  the source J1209$-$2032 from the VLBI FITS image database at 2.3, 5, and 7.6 GHz \citep{Beasley2002} reveal a pair of lobes, with total extension of $\sim$60 mas (300 pc). It is evident from the multi-frequency VLBI data that the western component contains the flat spectrum core. The core contributes nearly 56\% of the total flux at 5 GHz measured with the single dish Parkes telescope \citep[PKS90,][]{Wright1990}. 
    In the VLBA images of the source J1549$+$5038 at 5 GHz \citep{Xu1995,Helmboldt2007} and at 2 and 8 GHz \citep{Fey2000}, the source appears resolved into a dominant flat-spectrum core and a fairly bright curved jet which is $\sim$20 mas (180 pc) long and extends towards the south-west. The jet is itself resolved into a couple of knots. The core contributes nearly 67\% of the total flux at 5 GHz as measured with the single dish Green Bank telescope \citep[87GB,][]{Gregory1991}.
\end{itemize}

\begin{itemize}
    \item \textit{Optical Spectra}: \\
    J1209$-$2032 (z=0.404$\pm$0.002) is classified as a red elliptical galaxy
    with no sign of interaction with the nearby galaxies \citep{Jackson2002}. Its low-resolution spectrum obtained with the ESO 3.6m ESO reveals an early-type galaxy with old stellar population and weak emission lines [OII, OIII], indicating an AGN\citep{Hook2003}. There is no sign of variability seen in the two SDSS spectra taken 11 years apart. J1549$+$5038 ($z$=2.17428$\pm$0.00015) is a type 1 quasar with broad emission lines seen in its SDSS spectrum. No sign of merger or interaction is visible in the SDSS image.
\end{itemize}

As inferred from the inverted low-frequency radio spectra of both these sources, and also seen from the linear scales of their VLBI images, sources are compact, like GPS sources. But unlike the rolling down of the radio spectrum beyond the spectral peak, which is characteristic of GPS sources, the spectra of these two sources appear relatively flat at higher frequencies ($>$1 GHz)\citep{Mhaskey2019a,Mhaskey2019b}. Sparse and non-simultaneous spectral coverage could often lead to genuine GPS sources being misclassified as flat-spectrum quasars, and vice-versa \citep{Massaro2014}. Hence, J1209$-$2032 and J1549$+$5038 too could be of GPS type. 

\begin{table*}
\small
\addtolength{\tabcolsep}{-1pt}
\caption{The observational parameters of J1209$-$2032 and J1549$+$5038.}
\label{table:spec-prop}
\vspace{-0.5cm}
\begin{center}
\begin{tabular}{ccccccccc}\\
\hline\hline
\multicolumn{1}{c}{Coordinates}& \multicolumn{1}{c}{Opt}&$z$ &\multicolumn{1}{c}{150 MHz} & \multicolumn{1}{c}{325 MHz} & \multicolumn{1}{c}{1.4 GHz}&\multicolumn{1}{c}{Spectral Index} &\multicolumn{1}{c}{Power}& L$_{NUV}$\\
(J2000)&ID & &uGMRT$^{(a)}$ & uGMRT$^{a}$ & NVSS$^{(b)}$ & $\alpha$ & (1.4 GHz)& GALEX$^{(c)}$\\
& & &(mJy) &(mJy) & (mJy)& (150-325 MHz) & (W Hz$^{-1}$)& (W Hz$^{-1}$)\\
\hline
12 09 15.31 $-$20 32 34.4&G &0.404 &$<19.5$ & 149.4${\pm}$14.9 &353.7${\pm}$10.6$^{\ast}$&$>$2.64& 1.94$\times$10$^{26}$& $-$\\
15 49 17.46 +50 38 05.3&Q &2.17428&44.3${\pm}$6.2 & 287.8${\pm}$28.8 & 630.0${\pm}$18.9 &2.42${\pm}$0.22& 1.83$\times$10$^{28}$& 3.03$\times$10$^{23}$
\\
\hline
\end{tabular}

{$^{(a)}$ uGMRT -- \citet{Gupta2017,Mhaskey2019a,Mhaskey2019b}, 
$^{(b)}$ NVSS -- \citet{Condon1998}, $^{(c)}$ GALEX -- \citet{Bianchi2017}}
\end{center}
\end{table*}
%\end{landscape}
%\clearpage

\section{The uGMRT HI Observations and data analysis}
We observed the sources in our sample on 05 November 2019 using Band-5 (J1209$-$2032) and Band-3 (J1549$+$5038) of uGMRT.  For both the observations, a bandwidth of 12.5\,MHz split into 4096 frequency channels was used.  For J1209-2039, the observing band was centered at 1011.7\,MHz, and for J1549$+$5038 at 447.5\,MHz.
We observed the standard flux density calibrator 3C286 for 10\,mins at the beginning and at the end of each observing run to obtain a reliable flux density and bandpass calibrations. A nearby compact radio source was also observed periodically for phase calibration. The total on-source time for J1209-2039 and J1549$+$5038 was 70\,mins and 64\,mins, respectively. Only parallel polarisation products were recorded.  

The uGMRT data were processed using the Automated Radio Telescope Imaging Pipeline ({\tt ARTIP}) that is being developed to perform the end-to-end processing (i.e. from the ingestion of the raw visibility data to the spectral line imaging) of data from the uGMRT and MeerKAT absorption line surveys \citep{Gupta2016}. The pipeline is written using standard {\tt PYTHON} libraries and the {\tt CASA} package. The details are provided in \citet{Gupta2020}.  

In short, following data ingestion, the pipeline automatically identifies bad antennas, baselines, time ranges and radio frequency interference (RFI), using directional and median absolute deviation statistics.  After excluding these bad data, the complex antenna gains as a function of time and frequency are determined using the standard flux/bandpass and phase calibrators. Applying these gains, a continuum map that uses a user-defined range of frequency channels (excluding channels with absorption here) is made. Using this map as an input model, self-calibration complex gains are determined and then applied to all the frequency channels. Here, we performed 3 rounds of phase-only and 1 round of amplitude-and-phase self-calibration, and for imaging used {\tt ROBUST=0.5} weighting in {\tt tclean} task of {\tt CASA}.

\begin{figure}[hbt!]
\centering	
\includegraphics[trim={0cm 2cm 0cm 0cm},clip,width=\columnwidth]{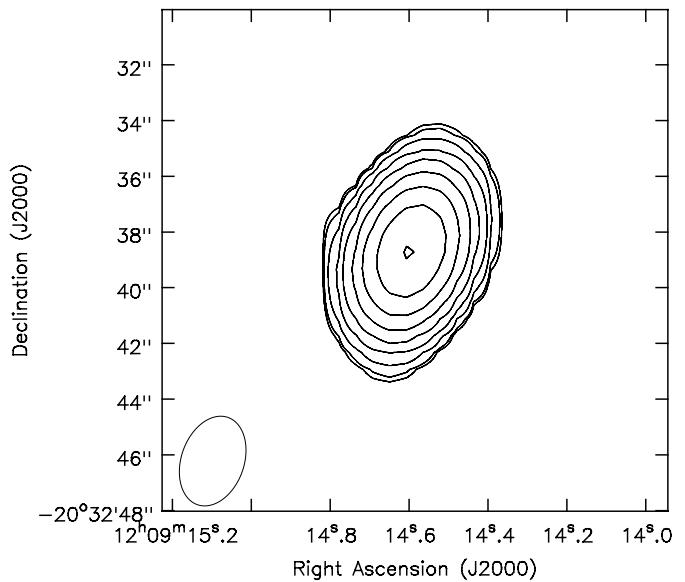}
\caption{uGMRT stokes-I image of the source J1209$-$2032 at 1011.7\,MHz. The contours are drawn at 3,4,8,16,32,64,128,256,512,1024,2048 times the
image rms noise which is 0.8 mJy. The beam size is $4.3^{\prime\prime}\times2.7^{\prime\prime}$ (position angle= -35$^\circ$).} 
\label{fig:HIJ1209-cont}  
\end{figure}

The radio source J1209$-$2032 is seen to be compact in our stokes-$I$ uGMRT image at 1011.7\,MHz (Figure~\ref{fig:HIJ1209-cont}). The image has a synthesized beam of $4.3^{\prime\prime}\times2.7^{\prime\prime}$ with a position angle of -35$^\circ$ and an rms of 0.8\,mJy\,beam$^{-1}$.  The deconvolved size of source is $0.8^{\prime\prime}\times0.5^{\prime\prime}$ (position angle = 125$^\circ$), consistent with the source being unresolved. The total flux density is 448.2\,mJy. 

The rms noise in the continuum image of J1549$+$5038 at 447.5\,MHz with a synthesized of $7.7^{\prime\prime}\times5.6^{\prime\prime}$ (position angle = 57$^\circ$) is 1.1\,mJy\,beam$^{-1}$.  The deconvolved size is $3.5^{\prime\prime}\times0.4^{\prime\prime}$ (position angle = 50$^\circ$). The source may be marginally resolved.  Its peak and total flux densities are 429\,mJy\,beam$^{-1}$ and 472\,mJy, respectively.

The {\tt ARTIP} then used CLEAN component models based on the above mentioned radio continuum maps to subtract the continuum emission from the {\it uv} data. This continuum-subtracted data set was then imaged to obtain the spectral-line cubes.

\begin{figure}[hbt!]
\centering	
\includegraphics[trim={0cm 1cm 0cm 2cm},clip,width=\columnwidth]{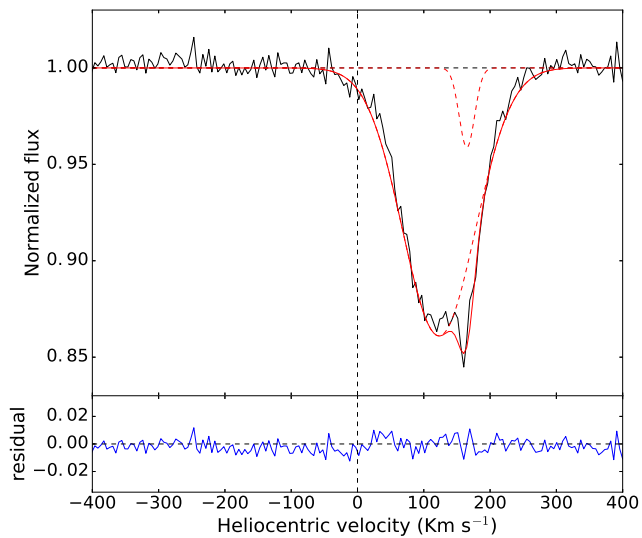}
\caption{Stokes-$I$ 21-cm absorption spectra towards J1209$-$2032. The spectrum has been smoothed over five consecutive spectral channels to improve the signal-to-noise ratio and normalized with respect to the peak flux density. 
The spectral resolution and rms noise in the smoothed spectra are 0.9\,\kms\ and 1.9\,mJy\,beam$^{-1}$\,channel$^{-1}$ respectively.}
\label{fig:HIJ1209}  
\end{figure}

\begin{figure}[hbt!]
\centering	
\includegraphics[trim={0cm 1cm 0cm 1cm},clip,width=\columnwidth]{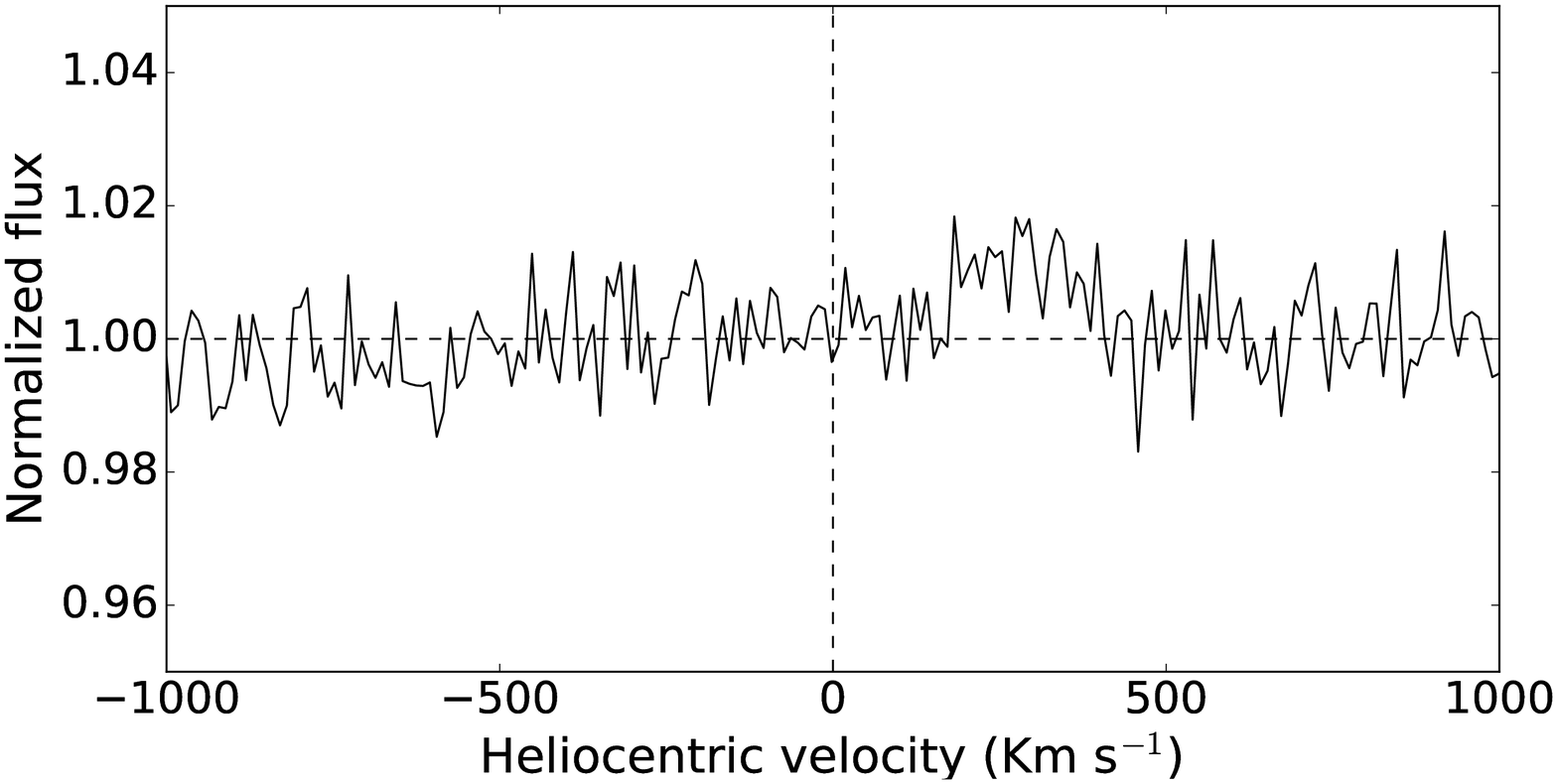}
\caption{Stokes-$I$ 21-cm absorption spectra towards J1549$+$5038. The spectrum has been smoothed over five consecutive spectral channels and normalized with respect to the peak flux density. The spectral resolution and rms noise in the smoothed spectra are 2.0\,\kms\ and 3.1\,mJy\,beam$^{-1}$\,channel$^{-1}$, respectively. The feature at 200\,\kms\ that looks like emission is due to weak RFI which is present only in LL.}
\label{fig:HIJ1549}  
\end{figure}

In Figures~\ref{fig:HIJ1209} and~\ref{fig:HIJ1549}, we present stokes-$I$ 21-cm absorption spectra towards the sources. The spectra have been smoothed over five consecutive spectral channels to improve the signal-to-noise ratio.  The spectral resolution and rms noise in the smoothed spectra are 0.9\,\kms\ and 1.9\,mJy\,beam$^{-1}$\,channel$^{-1}$ (J1209$-$2032), and 2.0\,\kms\ and 3.1\,mJy\,beam$^{-1}$\,channel$^{-1}$  (J1549$+$5038), respectively. The spectra have been normalized with respect to the peak flux density.
An absorption line is clearly detected towards J1209$-$2032 but not towards J1549$+$5038. The feature at 200\,\kms\ (Figure~\ref{fig:HIJ1549}) that looks like emission is due to weak RFI which is present only in LL.

For an optically thin cloud the peak 21-cm optical depth, $\tau_{p}$ is related to the neutral hydrogen column density

\begin{equation}
	N{(\HI)}=1.93\times10^{18}~\;\tau_{p}\;\frac{T_{s}}{f_{c}}\;\Delta v\; \textrm{cm}^{-2}.
\label{eq21cm}
\end{equation}

where, $T_{s}$ is the system temperature, $f_{c}$ is the covering factor, the fraction of the background emission covered by the absorber and $\Delta v$ is the full width half-maximum (FWHM) of the absorbed line.
%\begin{equation}
%    \textrm{N(HI)}\;=\;1.93\;\times\;10^{18}\;\tau_{p}\;\frac{T_{s}}{f_{c}}\;\Delta v\; \textrm{cm}^{-2}
%\end{equation}

HI line is detected towards the source J1209$-$2032, with a peak optical depth $\tau_{p}$=0.178$\pm$0.02 and a neutral hydrogen column density, N(HI)=34.8$\pm$2.9 $\times$10$^{20}$ cm$^{-2}$, assuming f$_c$ $=$ 1 and T$_s$ $=$ 100 K. Here we assume that the gas covers the radio source completely. However, the observed radio size from VLBI may only be a lower limit to the actual radio extent.

HI absorption is not detected towards the source J1549$+$5038. Taking its flux density as 429 mJy, based on the present observations, we place a 3$\sigma$ upper limit of $<$ 0.008 on the optical depth from the spectra shown in Figure~\ref{fig:HIJ1549}. The corresponding upper limit on the column density is N$_{HI} <$ 1.4 $\times$ 10$^{20}$ cm$^{-2}$ (assuming f$_c$ $=$ 1 and T$_s$ $=$ 100 K). 

\section{Discussion}
Since the source J1209$-$2032 with detected HI absorption, shows a well-resolved mildly asymmetric HI absorption line profile, it has been fitted with two Gaussian components. 
The velocity offset of the HI line centroid from the optical redshift is 144.01 kms$^{-1}$. For the first gaussian the peak optical depth and FWHM are 0.0437$\pm$0.015 and 32.128$\pm$5.74 kms$^{-1}$, respectively. The derived column density is N$_{HI}$= 2.7$\pm$1.1 $\times$ 10$^{20}$ cm$^{-2}$ for f$_c$ $=$ 1 and T$_s$ $=$ 100 K. For the second gaussian the peak optical depth and FWHM are 0.1345$\pm$0.0079 and 121.813$\pm$4.47 kms$^{-1}$ respectively. The derived column density is N$_{HI}$= 32.1$\pm$2.7 $\times$ 10$^{20}$ cm$^{-2}$ for f$_c$ $=$ 1 and T$_s$ $=$ 100 K. 

The HI profile is mildly shifted with respect to the systemic frequency. But, an error of 0.002 in redshift (z=0.404) of the source corresponds to an error of 1 MHz in its systemic frequency, which is approximately $\pm$300kms$^{-1}$ in the velocity range. Hence the observed shift is well within the error limits of the systemic frequency.    

Moreover, in general, velocities of $<\sim$200 kms$^{-1}$ at the systemic velocity can be just due to regular rotation in the galactic circumnuclear HI disk \citep{Pihlstrom2003}. The broad width observed in this case is due to rotational velocities, typically associated with the rotating gas structures in the host galaxies \citep{Morganti2018}. On the other hand, the line profiles with FWHM widths broader than $>$ 500 kms$^{-1}$ have to involve physical processes other than just rotational velocities, e.g., disturbed kinematics due to mergers, outflows, in order to accelerate the gas to such high velocities \citep{Gereb2015}.

Remarkably, the optical depth estimated for the source J1209$-$2032 is one of the highest known for GPS/CSS sources.  In a HI absorption study of galaxies with $z$ $>$ 0.2, \citet{Dutta2018} report that 40\% of sources with column density, HI $>$ 10$^{21}$ cm$^{-2}$ and 100\% of sources with HI $>$ 10$^{22}$ cm$^{-2}$ are associated with mergers. Also, the HI detection rate is high ($\sim$80\%) in merging systems. \citet{Dutta2018} used SDSS images of galaxies to identify signatures of galaxy mergers. However, no evidence of merger is found in the SDSS optical image 
of J1209$-$2032 (Section 2.1). 

The integrated optical depth can be used to derive a lower limit to the total gas content in the galaxy \citep{Pihlstrom2003}. The high value of optical depth suggests that the radio source J1209$-$2032, is embedded in a rich ISM, which has implications for the role of AGN feedback in the galaxy evolution \citep{Morganti2018}. Also, the dense environment indicated by the high optical depth would be consistent with the possibility of FFA mechanism being responsible for the sharp spectral turnover witnessed for the source J1209$-$2032, as inferred for GPS/CSS sources
\citep{Pihlstrom2003,Bicknell2018}. A discussion on the FFA process operating in 
such extremely inverted sources is provided in \citet{Mhaskey2019b}.

\subsection{Discussing the case of J1549$+$5038}
\subsubsection{The non-detection of HI absorption}
In the source J1549$+$5038, the non-detection of HI 21-cm absorption could be attributed to it being a quasar. The high redshift ($z$ = 2.171) of this source implies a high UV luminosity of $\sim$3$\times$10$^{23}$ W Hz$^{-1}$  at the NUV waveband ($\lambda_{eff}$ $\sim$ 2267\AA) in the GALEX observations \citep{Bianchi2017}. As argued by \citet{Curran2008}, at a UV luminosity above L$_{UV}$ $\sim$10$^{23}$W Hz$^{-1}$, the excitation and ionisation of the hydrogen gas may be severe, depleting the neutral hydrogen and thereby depressing HI absorption \citep{Curran2008,Curran2013,Morganti2018}. This argument was first advanced by \citet{Curran2008} to explain the striking lack of HI  absorption in the radio spectra of $z$ $>$ 2 \citep[see also,][]{Glowacki2017}. The UV luminosity of J1549$+$5038 is also close to the limit where all of the gas is expected to be ionised \citep{Curran+Whitting2012}. It is interesting that consistent with this suggestion, the most distant GPS radio source with detected HI absorption has only a modest redshift \citep[$z$ = 1.275,][]{Aditya2018}.

\subsubsection{Implications of a high RM}
Quite a few of the GPS/CSS sources are known to have high RM values ($>$1000 rad m$^{-2}$ in the rest frame). \citet{O'dea1998} compile a list of source of GPS/CSS sources with known RMs from \citet{Kato1987, Aizu1990, Taylor1992, Inoue1995}. Out of the 31 sources, 11 have RMs$>$1000 rad m$^{-2}$. Also, compared to galaxies, GPS sources associated with quasars have a wider range in RM and the smallest and the largest values of RM are identified with quasars. \citet{O'dea1998} also estimate the electron densities in the environment of GPS/CSS sources with large RMs. They find that the sources with high RMs are associated with high electron densities, supporting the notion that the large RMs in some GPS/CSS sources could be due to a dense magneto-ionic medium surrounding the source. Recently, broadband polarisation studies of a sample of compact AGNs by \citep{Pasetto2016}, with sources selected on the criterion of a strong depolarisation at 1.4 GHz, have revealed that the sources with young or newly ejected radio components (e.g. GPS/HFP sources) have a high rest-frame RM ($>$1000 rad $m^{-2}$). These sources are thus characterized by a dense magnetised medium that strongly rotates the angle of polarisation, leading to a large RM \citep{Pasetto2016}.

Thus, a high RM (1400$\pm$500 rad m$^{-2}$) in the source J1549$+$5038 indicates the presence of a dense magneto-ionic plasma. This provides a more direct way of estimating the ionised gas density in the environment of the source and supports the role of the in-homogeneous FFA model for the steep radio spectral turnover.

\subsection{Comparison with GPS/CSS sources}
Due to the compact nature of J1209$-$2032 and J1549$+$5038 and for the sake of uniformity with other EISERS which show a more convex radio spectrum, it is appropriate to compare the HI properties of these two sources with those of GPS/CSS sources.
In GPS/CSS sources, the HI absorption spectra more often indicates gas with disturbed kinematics traced by the asymmetric and highly blueshifted profiles \citep{Gereb2015,Glowacki2017,Morganti2018}. As discussed above, no such signature is evident in the HI spectrum of  the radio galaxy J1209$-$2032. One may note, however, that  in the sample of radio galaxies showing HI absorption taken from \citet{Gereb2015} and \citet{Glowacki2017}, only about 15\% show fast outflows with FWHM $>$ 500 km s$^{-1}$ \citep{Morganti2018}. Most of these outflows occur in either young radio galaxies or restarted radio galaxies \citep[][and references therein]{Aditya2018, Morganti2018}.

From a sample of 41 GPS/CSS sources compiled by \citet{Pihlstrom2003}, HI absorption is detected in 22 sources. In this sample, HI column densities of the sources are found to be anti-correlated with the radio source sizes. The optical depth was also found to be anti-correlated with source sizes, albeit with a large scatter \citep{Morganti2018} and, moreover, the HI column density is related to the linear size (L) as: N$_{HI}$ = 7.2 $\times$ 10$^{19}$ L$^{-0.43}$ cm$^{-2}$. \citet{Pihlstrom2003} suggest that this correlation is manifesting an ambient medium whose density decreases with radial distance from the centre. However, \citet{Curran2013} attributed this correlation to a geometrical effect, with a larger covering factor associated with smaller source sizes. Further, they propose that the relation is based on the assumption of a spin temperature and covering factor (where $T_{s}/f$ is 100. For sources where column densities are known, the $T_{s}/f$ is seen to vary by a factor of at least $\sim$170 (for $T_{s}$= 60K to 9950K). Therefore the above presumption and the observed large scatter in FWHM implies that the integrated optical depth is dominated by the peak optical depth, which is physically motivated and the observed relationship arises directly from an optical depth and linear size anti-correlation \citep{Curran2013}.

\begin{figure}[htb!]
\centering
\includegraphics[trim={0cm 3cm 0cm 5cm},clip,width=\columnwidth]{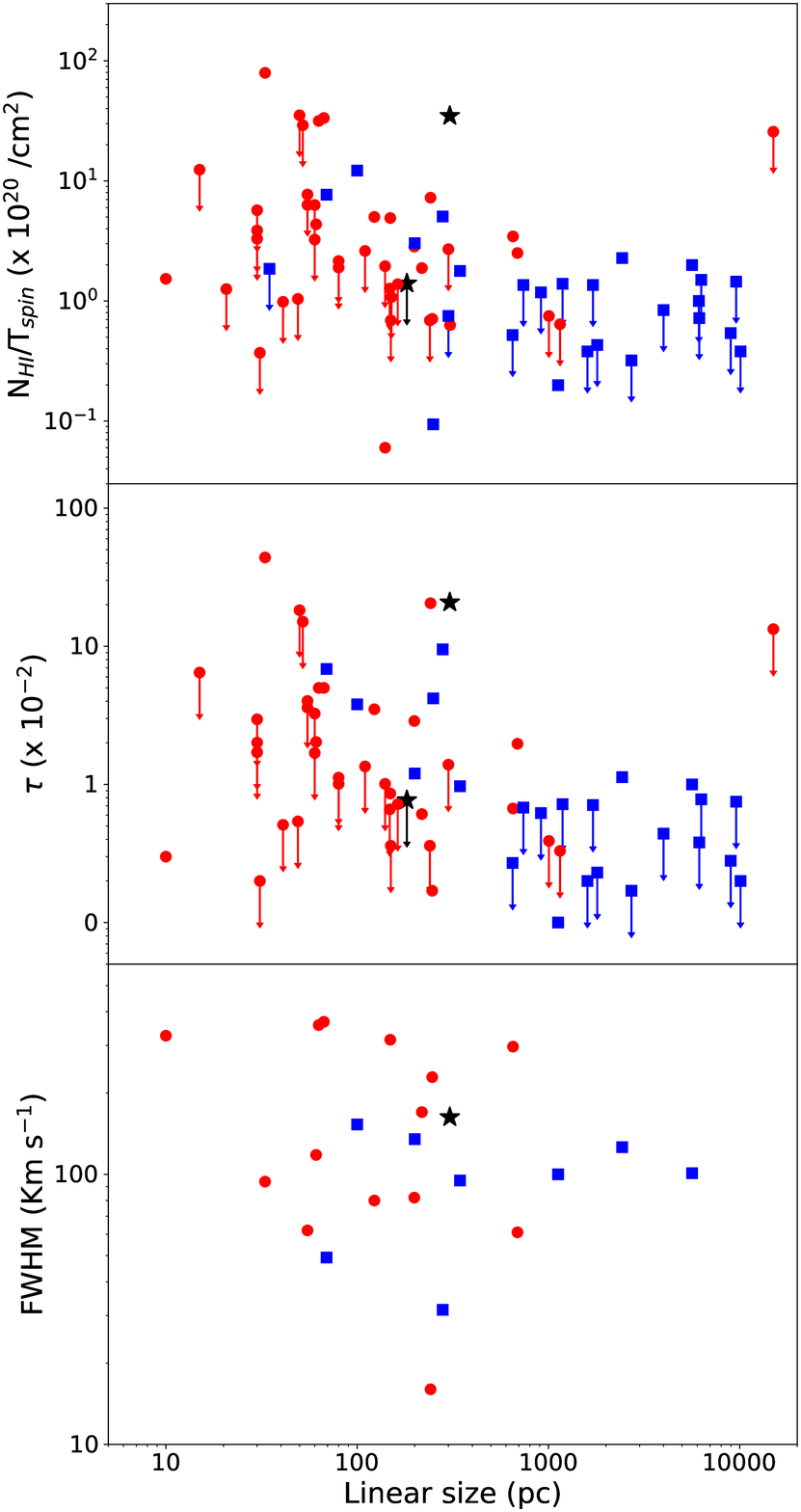}
\caption{Various spectral parameters for the present sample of GPS and CSS sources plotted against their linear sizes, with the superposition of the EISERS J1209$-$2032 and J1549$+$5038 (star symbol). The filled circles represent GPS sources and square symbol represents CSS sources. The errorbars for the source J1209$-$2032 are of the order of the symbol size.}
\label{fig:relations}  
\end{figure}

We further inspect this correlation using an extended and revised sample of GPS/CSS sources. The updated sample contains all the known GPS/CSS sources north of $-$30 degrees for which HI absorption has been searched, using various radio telescopes \citep{Vermeulen2003, Pihlstrom2003, Gupta2006,Chandola2011,Gereb2015,Grasha2019}. The GPS/CSS sources covered in these studies were drawn from: \citet{Spencer1989,Fanti1990,Sanghera1995,Taylor1996,deVries1997,Odea1997,Morganti1997,Peck2000,Xiang2005}. The sources classified as CSS sources in the literature, showing hints of spectral flattening towards lower radio frequencies (as seen in NED), but poorly sampled below the turnover frequency, were excluded. The final list consists of 48 GPS and 33 CSS sources. HI absorption is detected in 24 ($\sim$31\%) of the sources, as plotted in Figure~\ref{fig:relations} (see, Table~\ref{detection} and \ref{nondetection}). %(see Table A1 and A2; Appendix). 
We have also displayed in the plot our HI detection for the EISERS J1209$-$2032 and the upper limits for J1549$+$5038.

As seen from Figure~\ref{fig:relations}, the EISERS J1209$-$2032 does not follow the correlation between HI optical depth and linear size. The optical depth of J1209$-$2032 is high as compared to other GPS/CSS sources with similar linear sizes. This may be due to an asymmetric (non-spherical) distribution of the absorbing disk material relative to the radio structure, unlike the ideal spherical distribution assumed by \citet{Pihlstrom2003} or due to a geometrical effect as proposed by \citet{Curran2013} (see the discussion above).

%\begin{figure}[hbt!]

\section{Summary and Conclusions}
As an initial step towards characterising the occurrence and properties of the neutral gas in EISERS, we report here the 21cm spectral line study of EISERS. Using the uGMRT, we observed 2 out of the 7 known EISERS and detected HI absorption in the source J1209$-$2032 ($z$=0.404), identified with a galaxy. The detection of a very high optical depth (0.178$\pm$0.02) indicates a dense ISM conducive to the FFA scenario in this source. The moderate FWHM of the absorption profile (156$\pm$10.2 km/s), is typical of the rotating gas structures in galaxies. For the other source J1549$+$5049 ($z$=2.171), classified as quasar, we do not detect any HI absorption, which seems consistent with its high output of ionising UV photons. As the next step, we have planned HI spectroscopy of the remaining 5 EISERS, after obtaining their redshifts. This would allow us to place the EISERS reliably within the HI absorption distribution known for GPS/CSS sources, particularly in the context of the FFA scenario for the radio spectral turnover. 

%\clearpage  
\section{Acknowledgement}
We thank the staff of the GMRT who have made these observations possible. GMRT is run by the National Centre for Radio Astrophysics of the Tata Institute of Fundamental Research. This research has used NASA's Astrophysics Data System and NASA/IPAC Extragalactic Database (NED), Jet Propulsion Laboratory, California Institute of Technology under contract with National Aeronautics and Space Administration and VizieR catalogue access tool, CDS, Strasbourg, France. SP would like to thank DST INSPIRE Faculty Scheme (IF12/PH-44) for funding his research group. G-K acknowledges the Senior Scientist Fellowship of the Indian National Science Academy.

%\bibliography{references}

%\begin{thebibliography}
%\expandafter\ifx\csname natexlab\endcsname\relax\def\natexlab#1{#1}\fi
%\providecommand{\url}[1]{\href{#1}{#1}}

\clearpage

%\section{Appendix}
\begin{appendix} 
\begin{table*}[htb!]
\caption{List of GPS/CSS sources with HI detection used in Figure~\ref{fig:relations}, compiled from various sources in the literature.}\label{detection}
\centering
\small
\addtolength{\tabcolsep}{-4pt}
\begin{tabular}{ccccccccccccccc}
\hline 
S.no.	&	RA	&	 DEC	&	Name	&	Opt	&	$z$	&	Linear 	&	Radio 	&	Flux (mJy)	&	log P	&	N(HI) 	&	$\tau$ 	&	FWHM	&	Reference\\
 &		&		&		&	Class	&		&size (pc)	&Class	& 1.4 GHz	&1.4 GHz		&($\times$ 10$^{20}$ cm$^{-2}$)	&	$(\times 10^{-2}$)	&	(Kms$^{-1}$)	&	\\
\hline
1	&	00 25 49.1 	&	$-$26 02 13 	&	PKS 0023-26	&	 G	&	0.322	&	2429	&	CSS	&	8.75	&	30.484	&	2.28	&	1.13	&	126	&	1	\\
2	&	01 11 37.3 	&	+39 06 28 	&	J0111+3906	&	G	&	0.668	&	33	&	GPS	&	0.43	&	29.336	&	79.4	&	44	&	94	&	2,7	\\
3	&	01 19 35.0 	&	+32 10 50 	&	CGCG 502-027	&	 GPair	&	0.060	&	100	&	CSS &	2.64	&	29.383	&	12.2	&	3.8	&	153	&	3	\\
4	&	04 10 45.6 	&	+76 56 45 	&	4C +76.03	&	 G	&	0.599	&	689	&	GPS	&	5.62	&	30.434	&	2.51	&	1.97	&	61	&	1,7	\\
5	&	04 31 03.7 	&	+20 37 34 	&	PKS 0428+20	&	 G	&	0.219	&	653	&	GPS	&	3.76	&	30.004	&	3.45	&	0.67	&	297	&	1	\\
6	&	05 03 21.2 	&	+02 03 05  	&	J0503+0203	&	QSO	&	0.585	&	55	&	GPS	&	2.24	&	30.031	&	6.31	&	3.6	&	62	&	2,7	\\
7	&	07 57 56.7	&	+39 59 36	&	B3 0754+401	&		&	0.066	&	250	&	CSS	&	0.10	&	27.996	&	0.094	&	4.2	&		&	4	\\
8	&	11 24 43.7 	&	+19 19 28 	&	3C 258	&	 G	&	0.165	&	280	&	CSS	&	0.87	&	29.278	&	5.05	&	9.5	&	31.5	&	3	\\
9	&	11 50 00.1 	&	+55 28 21 	&		&	 G	&	0.139	&	93.9	&	CSS	&	0.15	&	28.460	&	6.31	&	10.5	&	42.1	&	5	\\
10	&	12 06 24.7 	&	+64 13 37 	&	3C 268.3	&	G	&	0.371	&	5641	&	CSS	&	3.72	&	30.149	&	1.99	&	1	&	101	&	2	\\
11	&	13 26 16.5 	&	+31 54 10 	&	4C +32.44	&	 G	&	0.368	&	247	&	GPS	&	4.86	&	30.264	&	0.71	&	0.17	&	229	&	1	\\
12	&	13 47 33.3 	&	+12 17 24 	&	4C +12.50	&	 G	&	0.122	&	200	&	CSS	&	5.40	&	29.962	&	3.03	&	1.2	&	135	&	3	\\
13	&	13 57 40.6 	&	+43 54 00 	&	B3 1355+441	&	 G	&	0.646	&	67	&	GPS	&	0.69	&	29.536	&	33.4	&	5	&	367	&	1	\\
14	&	14 00 28.6 	&	+62 10 39 	&	4C +62.22	&	 G	&	0.431	&	218	&	GPS	&	4.31	&	30.250	&	1.88	&	0.61	&	170	&	1	\\
15	&	14 07 00.4 	&	+28 27 15 	&	MRK 0668	&	 QSO	&	0.077	&	10	&	GPS	&	0.82	&	28.968	&	1.53	&	0.3	&	326	&	3	\\
16	&	14 09 42.4 	&	+36 04 16 	&		&	 G	&	0.148	&	69.2	&	CSS	&	0.14	&	28.457	&	7.67	&	6.85	&	49.2	&	5	\\
17	&	16 02 46.4 	&	+52 43 58 	&	4C +52.37	&	 G	&	0.106	&	345	&	CSS	&	0.57	&	28.933	&	1.78	&	0.97	&	94.9	&	5	\\
18	&	18 15 36.8 	&	+61 27 12 	&		&	 QSO	&	0.601	&	61	&	GPS	&	0.85	&	29.614	&	4.35	&	2.03	&	118	&	1,7	\\
19	&	18 16 23.9 	&	+34 57 46 	&	J1816+3457	&	G	&	0.245	&	123	&	CSS	&	0.73	&	29.324	&	5.01	&	3.5	&	80	&	2	\\
20	&	19 39 25.0	&	$-$63 42 46	&	PKS 1934$-$638	&	G	&	0.181	&	140	&	GPS	&		&		&	0.06	&		&		&	6	\\
21	&	19 44 31.5 	&	+54 48 07 	&		&	 G	&	0.263	&	149	&	GPS	&	1.75	&	29.728	&	4.91	&	0.86	&	315	&	1	\\
22	&	19 45 53.5 	&	+70 55 49 	&	J1945+7055	&	G	&	0.101	&	63	&	GPS	&	0.95	&	29.140	&	31.6	&	5	&	357	&	2	\\
23	&	20 52 52.0 	&	+36 35 35 	&	B2 2050+36	&	 G	&	0.354	&	242	&	GPS	&	5.14	&	30.278	&	7.25	&	20.51	&	16	&	1	\\
24	&	23 55 09.4 	&	+49 50 08 	&		&	 G	&	0.238	&	199	&	GPS	&	2.31	&	29.817	&	2.84	&	2.88	&	82	&	1	\\

\hline
\end{tabular}

{Reference: (1) \citet{Vermeulen2003}, (2) \citet{Pihlstrom2003}, (3) \citet{Gupta2006}, (4) \citet{Gereb2015}, (5) \citet{Chandola2011}, (6) \citet{veron2000}, (7) \citet{Grasha2019}  }\\
\end{table*}

\begin{table*}
\caption{List of GPS/CSS sources with no detected HI detection used in Figure~\ref{fig:relations}, compiled from various sources in the literature.}\label{nondetection}
\centering
\small
\addtolength{\tabcolsep}{-4pt}
\begin{tabular}{ccccccccccc}
\hline 
S.no.	&	RA	&	 DEC	&	Name	&	Opt	& $z$ &	Linear 	&	Radio 	&$\tau$ 	& N(HI) 	&	Reference$^{\dagger}$\\
 &		&		&		&	Class	&		&size (pc)&Morphology	&($\times 10^{-2}$)	& ($\times$ 10$^{20}$ cm$^{-2}$)	&	 \\
\hline
1	&	00 21 27.4	&	+73 12 41	&	0018+729	&		&	0.821  &140		&	GPS	&	$<$1.01	&	$<$1.95	&	5	\\
2	&	01 37 41.3 	&	+33 09 35 	&	3C 048	&	 QSO	&	0.367	&	6120	&	CSS	&		&	$<$1.00	&	1	\\
3	&	02 03 46.7	&	+11 34 45	&	PKS 0201+113	&		& 3.639	&	15	& GPS	&	$<$6.46	&	$<$12.4	&	5	\\
4	&	02 24 12.3 	&	+27 50 12 	&	3C 067	&	 G	&	0.310	&	9585	&	CSS	&	$<$0.75	&	$<$1.45	&	2	\\
5	&	02 40 08.2	&	$-$23 09 15	&	PKS 0237-23	&		&	2.223	& 150	&		GPS	&	$<$0.36	&	$<$0.69	&	5	\\
6	&	02 51 34.5 	&	+43 15 16 	&	[HB89] 0248+430	&	 QSO	&	1.310	&	110	&	GPS	& $<$1.35	&	$<$2.61	&	1	\\
7	&	02 55 39.2 	&	$-$21 53 52 	&	PKS 0253-220	&	 G	&	0.113	&	$<$4000	&	CSS	&	$<$1.91	&	<3.69	&	1	\\
8	&	03 01 42.4 	&	+35 12 21 	&	NGC 1167	&	 G	&	0.016	&	300	&	CSS	&		&	$<$0.75	&	1	\\
9	&	03 47 01.5 	&	$-$29 00 28 	&	PKS 0344-291	&	 GPair	&	0.142	&	<12500	&	CSS	&	$<$0.22	&	$<$0.42	&	1	\\
10	&	03 48 46.9 	&	+33 53 15 	&	3C 093.1	&	 G	&	0.243	&	1186	&	CSS	&	$<$0.72	&	$<$1.39	&	2	\\
11	&	04 32 36.5	&	+41 38 28	&	3C119	&		&	1.023	&650	& CSS	&	$<$0.27	&	$<$0.52	&	5	\\
12	&	04 59 52.1	&	+02 29 31	&	PKS 0457+024	&		&	2.384	& 55	&		GPS	&	$<$4.02	&	$<$7.7	&	5	\\
13	&	05 21 09.9 	&	+16 38 22 	&	3C 138	&	 QSO	&	0.759	&		&	CSS	&	$<$0.26	&	$<$0.49	&	2	\\
14	&	05 30 08.0	&	$-$25 03 29	&	PKS 0528-250	&		&	2.813	& 80	&	GPS	&	$<$	1.12	&	$<$	2.15	&	5	\\
15	&	05 42 36.1 	&	+49 51 07 	&	3C 147	&	 QSO	&	0.545	&	2717	&	CSS	&	$<$0.17	&	$<$0.32	&	2	\\
16	&	05 55 30.8	&	+39 48 49	&	B3 0552+398	&		&	2.365	& 30	&	GPS	&	$<$	2.96	&	$<$	5.7	&	5	\\
17	&	05 56 52.6 	&	$-$02 41 06 	&	PKS 0554-026	&	 G	&	0.235	&		&	GPS	&	$<$3.26	&	$<$6.28	&	2	\\
18	&	06 42 4.3	&	+67 58 35	&	HB89 0636+680	&		&	3.18	& 50	&	GPS	&	$<$	18.25	&	$<$	35.2	&	5	\\
19	&	07 33 28.6 	&	+56 05 42 	&		&	 G	&	0.104	&	151	&	GPS	&		&	$<$1.076	&	3	\\
20	&	07 41 10.7 	&	+31 12 00 	&		&	 QSO	&	0.632	&	41	&	GPS	&	$<$0.51	&	$<$0.984	&	2	\\
21	&	07 45 33.1	&	+10 11 12	&	PKS 0742+10	&		&	2.624	& 80	&	GPS	&	$<$	1.01	&	$<$	1.9	&	5	\\
22	&	07 53 03.3	&	+42 31 30 &	COINS J0753+4231	&		& 3.5892	&	60	&	GPS	&	$<$	3.26	&	$<$	6.3	&	5	\\
23	&	08 31 39.8 	&	+46 08 01 	&		&	 G	&	0.131	&	20.7	&	GPS	&		&	$<$1.256&	3	\\
24	&	09 00 40.0	&	$-$28 08 20	&	PKS 0858-279	&		&	2.152	& -	&	GPS	&	$<$	4.85	&	$<$	9.4	&	5	\\
25	&	09 36 09.4	&	$-$33 13 08	&		&		&	0.076	&		&	GPS	&		&	$<$0.055	&	4	\\
26	&	09 43 36.9 	&	$-$08 19 31 	&	PKS 0941-08	&	 G	&	0.228	&	148	&	GPS	&	$<$0.66	&	$<$1.27	&	2	\\
27	&	10 08 00.0	&	+07 30 16	&	3C237	&		&	0.87	& 10100	&		CSS	&	$<$	0.20	&	$<$	0.38	&	5	\\
28	&	10 35 07.0 	&	+56 28 47 	&		&	 G	&	0.459	&	163	&	GPS	&	$<$0.72	&	$<$1.38	&	2	\\
29	&	11 20 27.8 	&	+14 20 55 	&	4C +14.41	&	 G	&	0.362	&		&	GPS	&	$<$0.32	&	$<$0.61	&	2	\\
30	&	12 09 02.8 	&	+41 15 59 	&	B3 1206+415	&	 G	&	0.095	&	34.8	&	CSS	&		&	$<$1.854	&	3	\\
31	&	12 44 49.2	&	+40 48 06.2	&	COINS J1244+4048	&		&	0.813	& 300	& CSS	&	$<$	1.39	&	$<$	2.7	&	5	\\
32	&	12 52 26.3 	&	+56 34 20 	&	SBS 1250+568	&	 QSO	&	0.320	&	6146	&	CSS	&	$<$0.38	&	$<$0.72	&	2	\\
33	&	13 08 39.1 	&	$-$09 50 33 	&	PKS 1306-09	&	 G	&	0.467	&	1710	&	CSS	&	$<$0.71	&	$<$1.36	&	2	\\
34	&	13 35 22.0	&	+45 42 38	&	HB89 1333+459	&		&	2.449	& 15000	&	GPS	&	$<$	13.32	&	$<$	25.7	&	5	\\
35	&	14 43 14.5 	&	+77 07 28 	&	3C 303.1	&	 G	&	0.270	&	6295	&	CSS	&	$<$0.78	&	$<$1.5	&	2	\\
36	&	15 20 05.4	&	+20 16 05	&	3C318	&		&	1.574	& 8980	&		CSS	&	$<$	0.28	&	$<$	0.54	&	5	\\
37	&	15 21 14.4 	&	+04 30 22 	&	4C +04.51	&	 G	&	1.296	&	1150	&	GPS	&	$<$0.33	&	$<$0.64	&	1	\\
38	&	15 46 09.5 	&	+00 26 25 	&	PKS 1543+005	&	 G	&	0.550	&	49	&	GPS	&	$<$0.54	&	$<$1.04	&	2	\\
39	&	16 04 01.5 	&	$-$22 23 41 	&	PKS 1601-222	&	 G	&	0.141	&	30	&	GPS	&	$<$1.71	&	$<$3.30	&	1	\\
40	&	16 34 33.8	&	+62 45 35	&	3C343	&		&	0.988	& 1600	& CSS	&	$<$	0.20	&	$<$	0.38	&	5	\\
41	&	16 38 28.2	&	+62 34 44	&	3C343.1	&		&	0.75	& 1800	& CSS	&	$<$	0.23	&	$<$	0.43	&	5	\\
42	&	17 03 30.4 	&	+45 40 47 	&	B3 1702+457	&	 G	&	0.060	&	$<$8.4	&	CSS	&		&	$<$4.165&	3	\\
43	&	18 23 14.1 	&	+79 38 49 	&		&	 QSO	&	0.224	&	52	&	GPS	&	$<$15.08	&	$<$29.1	&	2	\\
44	&	18 31 14.8 	&	+29 07 10 	&	4C +29.56	&	 G	&	0.842	&		&	CSS	&	$<$0.89	&	$<$1.71	&	2	\\
45	&	18 45 35.1 	&	+35 41 17 	&	LQAC 281+035 002	&	 G	&	0.764	&		&	GPS	&	$<$5.70	&	$<$11	&	2	\\
46	&	18 50 27.6	&	+28 25 13	&	TXS 1848+283	&		&	2.56	& -	&	GPS	&	$<$	10.86	&	$<$	20.9	&	5	\\
47	&	20 03 24.1	&	$-$32 51 45	&	PKS 2000-330	&		&	3.773	& -	&	GPS	&	$<$	19.13	&	$<$	36.9	&	5	\\
48	&	20 22 06.7 	&	+61 36 59 	&		&	Q	&	0.227	&	31	&	GPS	&	$<$0.2	&	$<$0.37	&	2	\\
49	&	20 58 28.8 	&	+05 42 51 	&	4C +05.78	&	 G	&	1.381	&	$<$3400	&	CSS	&	$<$0.74	&	$<$1.42	&	1	\\
50	&	21 29 12.2	&	$-$15 38 41	&	PKS 2126-15	&		&	3.268	& 60	&	GPS	&	$<$	1.69	&	$<$	3.25	&	5	\\
51	&	21 30 32.9	&	+05 02 17	&	PKS 2127+04	&		&	0.99	& 240	&	GPS	&	$<$	0.36	&	$<$	0.69	&	5	\\
52	&	21 37 50.0 	&	$-$20 42 32 	&	PKS 2135-20	&	 G	&	0.636	&	914	&	CSS	&	$<$0.62	&	$<$1.18	&	2	\\
53	&	21 51 37.9	&	+05 52 13	&	PKS 2149+056	&		&	0.74	& 30	&		GPS	&	$<$	2.01	&	$<$	3.87	&	5	\\
54	&	22 50 25.3 	&	+14 19 52 	&	[HB89] 2247+140	&	 QSO	&	0.235	&	740	&	CSS	&	$<$0.68	&	$<$1.36	&	2	\\
55	&	23 25 42.3 	&	+43 46 59 	&	B3 2323+435A	&	 G	&	0.145	&	4000	&	CSS	&	$<$0.44	&	$<$0.84	&	1	\\
56	&	23 44 03.8 	&	+82 26 40 	&	[HB89] 2342+821	&	 QSO	&	0.735	&	1006	&	GPS	&	$<$0.39	&	$<$0.75	&	2	\\

\hline
\end{tabular}

{$\dagger$Reference: (1) \citet{Gupta2006}, (2) \citet{Vermeulen2003}, (3) \citet{Chandola2011}, (4) \citet{Gereb2015}, (5) \citet{Grasha2019} }\\
\end{table*}
\end{appendix}

\end{document}